\documentclass[aps,prl,reprint,amsmath,amssymb,superscriptaddress]{revtex4-2}

\usepackage{color}
\usepackage[colorlinks=true,citecolor=magenta]{hyperref}

\usepackage{graphicx}
\usepackage{bm}
\usepackage{braket}		
\usepackage[normalem]{ulem}

\newcommand{\beq}{\begin{equation}}
\newcommand{\eeq}{\end{equation}}
\newcommand{\bea}{\begin{eqnarray}}
\newcommand{\eea}{\end{eqnarray}}
\def\be{\begin{equation}}
\def\ee{\end{equation}}

\newcommand{\mpl}{M_{\rm Pl}}

\renewcommand{\L}{\mathcal L}

\newcommand{\del}{\partial}

\newcommand{\cA}{\mathcal{A}}

\def\be{\begin{equation}}
\def\ee{\end{equation}}
\def\ba{\begin{eqnarray}}
\def\ea{\end{eqnarray}}

\def\p{\partial}
\def\d{\mathrm{d}}
\def\mn{_{\mu \nu}}

\def\({\left(}
\def\){\right)}

\def\b3{b_3}
\def\bt3{\tilde b_3}

\def\nn{\nonumber}

\def\d{\mathrm{d}}

\usepackage[normalem]{ulem}
\usepackage{xcolor}

\def\ba{\begin{eqnarray}}
\def\ea{\end{eqnarray}}

\begin{document}

\preprint{Imperial/TP/2021/LA/1}

\title{Reverse Bootstrapping: IR lessons for UV physics}


\author{Lasma Alberte}
\email[]{l.alberte@imperial.ac.uk}
\affiliation{Theoretical Physics, Blackett Laboratory, Imperial College, London, SW7 2AZ, U.K.}
\author{Claudia de Rham}
\email[]{c.de-rham@imperial.ac.uk}
\affiliation{Theoretical Physics, Blackett Laboratory, Imperial College, London, SW7 2AZ, U.K.}
\affiliation{CERCA, Department of Physics, Case Western Reserve University, 10900 Euclid Ave, Cleveland, OH 44106, USA}
\author{Sumer Jaitly}
\email[]{sumer.jaitly14@imperial.ac.uk}
\affiliation{Theoretical Physics, Blackett Laboratory, Imperial College, London, SW7 2AZ, U.K.}
\author{Andrew J. Tolley}
\email[]{a.tolley@imperial.ac.uk}
\affiliation{Theoretical Physics, Blackett Laboratory, Imperial College, London, SW7 2AZ, U.K.}
\affiliation{CERCA, Department of Physics, Case Western Reserve University, 10900 Euclid Ave, Cleveland, OH 44106, USA}

\date{\today}

\begin{abstract}
S-matrix bootstrap and positivity bounds are usually viewed as constraints on low-energy theories imposed by the requirement of a standard UV completion. By considering graviton--photon scattering in the Standard Model, we argue that the low-energy theory can be used to put constraints on the UV behaviour of the gravitational scattering amplitudes.
\end{abstract}


\maketitle

\textit{\textbf{Introduction.---}} In recent years S-matrix positivity bounds and bootstrap methods have matured into a powerful tool to constrain low-energy effective field theories (EFTs). The original nonlinear S-matrix positivity bounds of the 1970's were largely concerned with constraints on individual partial wave coefficients where often experimental data was forthcoming, or constraints on the amplitude within the Mandelstam triangle which is only well defined in gapped theories \cite{yndurain1972rigorous}. The modern use of positivity bounds was reinvigorated in \cite{Adams:2006sv}, based on earlier work \cite{Pham:1985cr,Ananthanarayan:1994hf}, where it was emphasized that these act as constraints on EFTs, including on massless ones once mild assumptions are made.
These constraints are interconnected with causality considerations and  for Lorentz invariant non-gravitational theories where causality and locality are precisely defined, there are now a large number of robust bounds on Wilson coefficients from EFTs. The linear forward limit bounds of  \cite{Adams:2006sv} were extended away from the forward limit in \cite{deRham:2017avq}, and for particles of arbitrary spins in \cite{Bellazzini:2016xrt,deRham:2017zjm}.
These generalized bounds have been used to provide powerful constraints on low-energy EFTs for instance in \cite{Cheung:2016wjt,Bonifacio:2016wcb,deRham:2017imi,deRham:2017xox,deRham:2018qqo,Afkhami-Jeddi:2018own,Zhang:2018shp,Bellazzini:2019bzh,Melville:2019wyy,Alberte:2019xfh,Alberte:2019zhd,Kim:2019wjo,Herrero-Valea:2019hde,Remmen:2019cyz,Remmen:2020uze,Wang:2020jxr,deRham:2021fpu,
Traykova:2021hbr,Davighi:2021osh,Bern:2021ppb}.
Subsequently the linear positivity bounds were generalized to a set of nonlinear constraints in \cite{Arkani-Hamed:2020blm,Chiang:2021ziz,Bellazzini:2020cot} using the same methodology as Stieltjes moment positivity bounds derived in \cite{common1969properties,yndurain1969constraints,common1970some,common1970constraints}. More recently, the full use of crossing symmetry has tightened these nonlinear statements \cite{Tolley:2020gtv,Caron-Huot:2020cmc,Sinha:2020win,Du:2021byy,Haldar:2021rri,Raman:2021pkf}, which strongly overlap with S-matrix bootstrap bounds \cite{Paulos:2017fhb,Guerrieri:2020bto,Hebbar:2020ukp,Guerrieri:2021ivu}.\\

The application of these methods to theories with gravity is however less developed since the precise rules for causality
are far less established \cite{Gao:2000ga,Camanho:2014apa,Hollowood:2015elj,deRham:2019ctd,deRham:2020zyh,Chen:2021bvg}. In  \cite{deRham:2020zyh,Alberte:2020jsk} it was argued that based on causality considerations, the usual amplitude combinations which are demanded to be positive could admit a small Planck scale suppressed negativity in the presence of gravity, and this was confirmed by the novel impact-parameter bounds of~\cite{Caron-Huot:2021rmr} as well as in \cite{Tokuda:2020mlf}. This apparent gravitational weakening of positivity bounds is intimately connected with the fact that perturbative corrections to the sound speed in gravitational EFTs can appear to be superluminal \cite{Drummond:1979pp,Lafrance:1994in,Shore:2002gw,Hollowood:2007kt,Hollowood:2007ku,Hollowood:2008kq,
Goon:2016une,deRham:2020ejn,Bellazzini:2021shn,AccettulliHuber:2020oou,deRham:2021bll}  even though causality is never violated as these effects are not resolvable \cite{Hollowood:2015elj,deRham:2019ctd,deRham:2020zyh,Chen:2021bvg}.\\

There are two central problems with extending the usual positivity methods to $2$--$2$ scattering amplitudes when including gravity:

The first is that
massless graviton loops give rise to a branch cut which extends to $t=0$, preventing the continuation of the partial wave expansion from $t<0$ to $t \ge 0$ where the positivity bounds ought to be satisfied (in terms of the standard $s,t,u$ Mandelstam variables).
 However since the start of the branch cut is associated to light loops, progress can be made by either directly removing the low-energy loops in the manner of the improved positivity bounds \cite{Bellazzini:2016xrt,deRham:2017imi}, or applying the bounds to the tree level of the massless and light states, while loops of the heavy fields can be included or integrated out (see, e.g \cite{Caron-Huot:2021rmr,Bern:2021ppb}). Since loops of massless modes are not included, this also removes any issue with IR divergences which are pertinent in four dimensions.

 The second  is the presence of a massless $t$-channel pole associated with gravitational exchange. Since this pole grows as $s^2$, it is not possible to subtract it and to continue to use a dispersion relation with two subtractions as is allowed for non-gravitational theories. However there is no difficulty in working with a dispersion relation with more than two subtractions and hence many non-trivial positivity bounds and S-matrix bootstrap constraints have been applied by focussing on the higher order EFT operators (see for example \cite{Bern:2021ppb} for an excellent recent discussion).

 The impact-parameter bounds of~\cite{Caron-Huot:2021rmr} evade both problems by working at $t<0$  and looking for a new set of positive integrals not immediately related to the standard bounds. The resulting bounds are then consistent with those conjectured in \cite{Alberte:2020jsk,Alberte:2020bdz} but  controlled by the lightest massive state integrated out.

One approach to tighten this bound  is to assume a Regge behaviour for the UV completion \cite{Hamada:2018dde,Tokuda:2020mlf,Herrero-Valea:2020wxz,Noumi:2021uuv,Aoki:2021ckh} as given below in Eq.~\eqref{eq:Regge}. This behaviour arises in weakly coupled string theory, but it can also be argued for universally, and folding this information into the bounds of the graviton-photon scattering either leads to the presence of new tower of higher spin states starting at least at the TeV scale, or a violation of the Froissart bound, an IR/UV mixing, or a constraint on the slope of the  residue of the Regge pole.\\

\textit{\textbf{Gravitational positivity bounds.---}} In what follows, we shall consider scattering amplitudes that can be consistently computed while including tree and loop level contributions from all massive states, but only trees from massless ones (i.e. no massless loops). At low energies, these amplitudes admit an expansion which is determined by the tree amplitudes of the low-energy EFT obtained from integrating out all massive states. In the specific context of graviton scattering with other light states $Xh\rightarrow Xh$, the dangerous graviton loops may be removed by taking a scaling limit of the exact scattering amplitude $\cA(s,t)$ as
\be
\label{eq:Atilda}
\tilde \cA_{Xh\rightarrow Xh}(s,t) \equiv \lim_{\mpl \rightarrow \infty} \mpl^2 \, \cA_{Xh\rightarrow Xh}(s,t) \, .
\ee
The central point is that graviton loops enter the amplitude $\cA$ only at order $1/\mpl^4$, and so by taking this scaling, graviton loops are automatically projected out in an RG independent manner, \footnote{Graviton loops can only contribute to the amplitude if higher order operators are tuned to enter at an extremely low cutoff scale. Such effects are for instance considered in section 4.4 of Ref.~\cite{Alberte:2020bdz}.}. This reduced amplitude satisfies unitarity in the sense \footnote{We define the amplitude discontinuity by ${\rm Disc}(\cA)= \frac{1}{2 i} \left(\cA(s+i \epsilon)-\cA(s-i \epsilon)\right)$}
\ba
{\rm Disc}\tilde \cA_{Xh\rightarrow Xh} = \frac{1}{2}\sum_Y \Big[ (2 \pi)^4 \delta^4(k_h+k_X-k_Y) \\
 \times \tilde \cA_{Xh \rightarrow Y} {\tilde \cA}^*_{Xh \rightarrow Y}\Big]\,,\nn
\ea
where  $Y$ is a complete set of states in the UV completion not including gravitons and
\be
 \tilde \cA_{Xh \rightarrow Y} \equiv \lim_{\mpl \rightarrow \infty} \mpl \, \cA_{Xh \rightarrow Y} \, ,
 \ee
which is enough to ensure positivity for elastic scattering processes. With the graviton loops removed, the only remaining dangerous singularities are the graviton $t$-channel pole and the loops of any other massless states such as the photon. For the process we shall be considering, the latter are largely harmless and will not contribute at the order we shall be interested in. \\

 We now make the standard assumption that the amplitude $\tilde \cA_{Xh\rightarrow Xh}(s,t) $ admits a dispersion relation with two subtractions in the physical region $t<0$. Although the Froissart bound \cite{Froissart:1961ux,PhysRev.135.B1375,Martin:1965jj} does not strictly apply to massless states, reasonable causality considerations applied to the scattering amplitude in impact parameter space in the physical region imply the bound
 \be
 \label{eq:Froissart}
\lim_{|s| \rightarrow \infty }\ s^{-2} \tilde \cA_{Xh\rightarrow Xh}(s,t) =0  \quad {\rm for }\quad t<0\,,
 \ee
throughout the complex $s$ plane.  In particular in $D \ge 5$ this bound is expected for low-energy EFTs which descend from string theory, and so a violation of the $<|s|^2$ growth would hence violate predictions from perturbative string theory.
In $D=4$ the situation is more subtle because of IR divergences, however by working in the scaling limit \eqref{eq:Atilda} we have removed any dangerous IR contributions from graviton loops. 
Hence assuming \eqref{eq:Froissart}, and assuming that light loops do not spoil the standard analyticity conditions,  the amplitude enjoys a twice-subtracted dispersion relation for $t<0$,
\ba\label{disp1}
&&\tilde \cA_s(s,t) = a_s(t) + b_s(t) s + \left(\text{$s$- and $u$-channel poles}\right)  \\
&&+ \frac{s^2}{\pi}\int_{0}^{\infty} \d \mu \frac{{\rm Disc}\tilde \cA_{s}(\mu,t)}{\mu^2(\mu-s)}+  \frac{u^2}{\pi}\int_{0}^{\infty} \d \mu \frac{{\rm Disc}\tilde \cA_{u}(\mu,t)}{\mu^2(\mu-u)} \, ,\nn
\ea
where $\tilde \cA_s$ denotes the $s$-channel process $Xh\rightarrow Xh$ and $\tilde\cA_u$ --- the crossed process $X \bar h \rightarrow X\bar h$. The key observation is that while the amplitude on the left-hand side of  \eqref{disp1} contains a spin-2 $t$-channel pole, the pole does not explicitly appear in the dispersion relation valid for $t<0$, on the right-hand side of \eqref{disp1}. Hence the pole is found within the dispersive integral. More concretely, as we approach $t=0$ from below,
\be
\lim_{t \rightarrow 0^-}\( \int_{0}^{\infty} \d \mu \frac{{\rm Disc}\,\tilde \cA_{s}(\mu,t)}{\mu^2(\mu-s)} +s\leftrightarrow u \) \sim \frac{1}{t}\,.
\ee
Since the pole does not arise in the discontinuity, it must arise from the failure of the integral to converge as $t \rightarrow 0$. At the same time we know that a dispersion relation with three subtractions is well-behaved even for $t>0$ (again assuming massless loops do not contribute). This implies that $\int_{0}^{\infty} \d \mu \frac{{\rm Disc}\,\tilde \cA_{s , u}(\mu,t)}{\mu^3(\mu-s)} $ is a convergent integral for small $t>0$. We thus conclude that as $\mu \rightarrow \infty$ the discontinuity behaves as
\ba
\begin{cases}
 {\rm Disc}\,\tilde \cA_{s , u}(\mu,t)<    \mu^2, & \text{for } t<0\,, \\
   \mu^2< {\rm Disc}\,\tilde \cA_{s , u}(\mu,t)<  \mu^3 , & \text{for small } t>0.
\end{cases}
\ea
Assuming the mildest analytic behaviour for the $t$-dependence of the discontinuity in either of the $s$- and $u$-channels, we are necessarily led to the Regge assumption for fixed $t$ near $t=0$ \cite{Hamada:2018dde,Tokuda:2020mlf,Herrero-Valea:2020wxz,Noumi:2021uuv,Aoki:2021ckh},
\be
\label{eq:Regge}
\lim_{\mu \rightarrow + \infty} {\rm Disc}\,\tilde \cA_{s,u}(\mu,t) = r_{s,u}(t) \Lambda_r^4 \( \frac{\mu}{\Lambda_r^2}\)^{\alpha_{s,u}(t)}\,,
\ee
where $\alpha(t)$ is the Regge trajectory, and $r(t)$ is related to the residue of the associated Regge pole. The scale $\Lambda_r$ is freely chosen as it can be absorbed into $r(t)$ and is introduced for later convenience.
The Regge slopes satisfy $\alpha(t)<2$ for $t<0$ and $\alpha(t)>2$ for $t>0$ \footnote{Often we assume the same Regge slope as implied by the Pomeranchuk theorem however this is not necessary for our argument.}.
Given the absence of massless loops we expect $\alpha(t)$ to be analytic at $t=0$.  We stress that we are led to this Regge assumption without any input from string theory, although the latter is certainly consistent with it (see e.g. \cite{Veneziano:1968yb,Collins:1977jy}). Defining for each channel the difference,
\be
R(\mu,t) \equiv {\rm Disc}\,\tilde \cA(\mu,t)-r(t) \Lambda_r^4 \( \frac{\mu}{\Lambda_r^2}\)^{\alpha(t)}\,,
\ee
(where we omit the $s,u$ subscripts unless needed),
the dispersion relation may be reorganized into
\ba\label{disp3}
&&\tilde \cA(s,t) = a_s(t) + b_s(t) s + \left(\text{$s$-channel poles}\right) \nn \\
&&+ \frac{s^2 r_s(t)}{\pi (2-\alpha_s(t))}+ \frac{s^2}{\pi}\int_{0}^{\Lambda_r^2} \d \mu \frac{{\rm Disc}\tilde \cA_{s}(\mu,t)}{\mu^3}  \\
&&+ \frac{s^2}{\pi}\int_{\Lambda_r^2}^{\infty} \d \mu \frac{R_s(\mu,t) }{\mu^3}+\frac{s^3}{\pi}\int_{0}^{\infty} \d \mu \frac{{\rm Disc}\tilde \cA_{s}(\mu,t)}{\mu^3(\mu-s)}+ {s \leftrightarrow u} \, . \nn
\ea
The crucial difference is that the dispersion relation \eqref{disp3} is now also valid for $t>0$, unlike \eqref{disp1}, since all of the dispersive integrals are convergent. The $t$-channel pole is now explicit in the Regge slope contribution
\be
\lim_{t \rightarrow 0} \frac{s^2 r(t)}{\pi (2-\alpha(t))} \sim -\frac{s^2 r}{\pi \alpha' t} \, ,
\ee
given that $\alpha(t)$ is analytic in $t$ at $t=0$, so that $\alpha(t)=2+\alpha't+ \mathcal{O}(t^2)$.  Thus defining the amplitude with poles removed in all three channels in a crossing symmetric way
\be
\hat \cA(s,t) \equiv \tilde \cA(s,t) - \(\text{$s$-, $u$- and $t$-channel poles}\)\,,
\ee
we infer the forward limit positivity bound
\ba
\partial_s^2 \hat \cA(0,0) &>& - \frac{r_s}{\pi \alpha_s'}  \( 2 \frac{r_s'}{r_s}-\frac{\alpha''_s}{{\alpha'_s}}\)  \nn \\
&+& \frac{2}{\pi}\int_{\Lambda_r^2}^{\infty} \d \mu \frac{R_{s}(\mu,0) }{\mu^3}+ {s \leftrightarrow u} \, .
\ea
Now crucially the above formula is valid for any value of $\Lambda_r$, and in particular we are free to choose $\Lambda_r$ to be some scale much larger than the actual scale at which the Regge behaviour kicks in. Since $R(s,t)$ is the subleading term in the Regge behaviour, we would expect it to be suppressed by more than $\ln s$ relative to the leading part~\footnote{A suppression in $R(\mu,t)/(r(t) \mu^{\alpha(t)})$ of only $\ln(\mu)$ would give rise to a $\log(t)$ branch cut which is expected from loops, however are not included in $\tilde{\mathcal{A}}$ by construction \cite{Herrero-Valea:2020wxz}.}. Hence we may scale $\Lambda_r \rightarrow \infty$ to ensure that the bound is effectively
\be\label{bound1}
\partial_s^2 \hat \cA(0,0) > - \beta_s \( 2 (\ln r_s)'-(\ln \alpha'_s)'\) + {s \leftrightarrow u} \, ,
\ee
where we have defined $\beta_{s,u} = r_{s,u} /(\pi \alpha'_{s,u})$. The $\beta$'s combined as $\beta=\beta_s+\beta_u$  which can be matched against the actual low energy $t$-channel pole
\be
\tilde \cA(s,t) = - \beta \frac{s^2}{ t}+\hat \cA(s,t) + \(\text{$s$ and $u$ poles}\)\, .
\ee
In practice for the scattering of only massless states, $\beta$ is either of order one or zero depending on whether the given process allows $t$-channel graviton exchange.
Moreover, the slope of the Regge residue is always positive by virtue of unitarity and the partial wave expansion
\be
\partial_t \( {\rm Disc}  \tilde \cA_{s , u}(\Lambda_r^2,0)\)> 0 \, \Rightarrow r_{s,u}'>0 \, .
\ee
In the rest of this work we shall focus on the photon--graviton scattering process $Ah\to Ah$ accounting for Standard Model (SM) effects and inferring the implications of the bound \eqref{bound1}. Remarkably we shall see that this provides us a bound on UV rather than on IR physics. \\

\begin{figure}[]
\centering
\includegraphics[width=\textwidth]{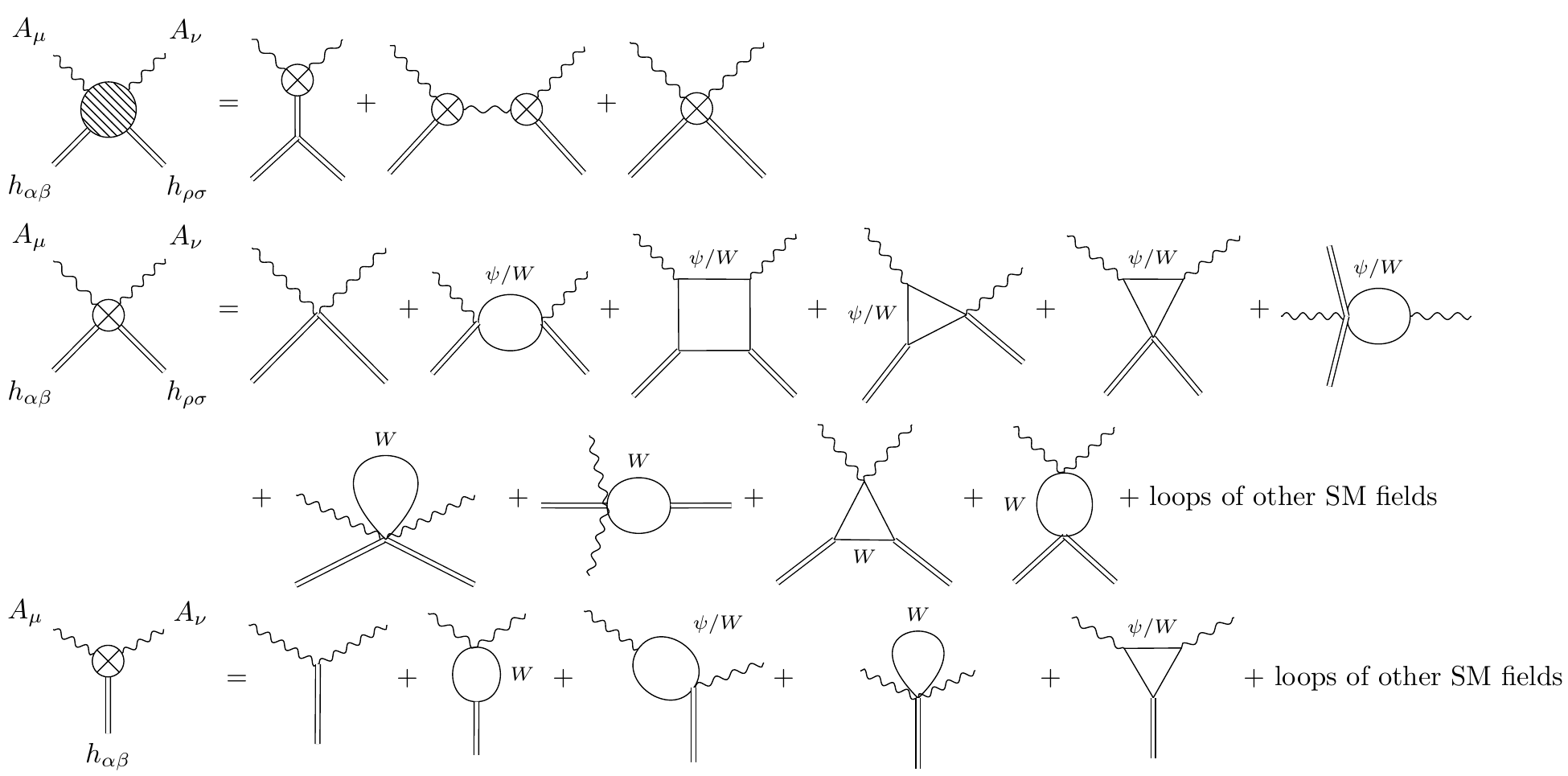}
\caption{Top line: contributions to the $Ah\to Ah$  scattering amplitude $\tilde{\mathcal{A}}$ defined in \eqref{eq:Atilda}  at order $s^2$. Second (resp. third) line:  redressed off-shell vertex $AAhh$ (resp. $AAh$) from electroweak loops. Hadronic loops can be computed in a similar way. The amplitude is dominated by loops from the lightest charged field.}\label{fig:diagrams}
\end{figure}

\textit{\textbf{$\bm{AhAh}$ Positivity.---}}
Due to the universal nature of the graviton coupling all electroweak and QCD sector particles contribute to the $Ah\to Ah$ scattering process. We may start by considering all the SM particles to be minimally coupled to gravity in a covariant way, although the implications of this work are insensitive to that assumption,
\ba\label{SM}
\L=-\frac{\mpl^2}{2}R+\L_{\rm SM}(g\mn,A_\mu, \psi, W^{\pm},Z,{\rm QCD},\cdots)\,.\quad
\ea
The graviton enters the metric as $g\mn=\eta\mn+2h\mn/\mpl$, and $A_\mu$ designates the photon. Every charged lepton $\psi$, the $W$ bosons and the QCD sector enter the photon--graviton scattering and the relevant diagrams are schematically shown as supplementary material in Fig.~\ref{fig:diagrams}. In practise however, up to order $s^2$ in the amplitude, the effects of all contributions from the SM to the $Ah\to Ah$ amplitude can be captured by the following operators
\ba\label{EH}
\L=-\frac{\mpl^2}{2}R-\frac14F_{\mu\nu}F^{\mu\nu}+ \b3  F_{\mu\nu}F_{\rho\sigma}R^{\mu\nu\rho\sigma}+\mathcal{O}_{\rm dim\ge8}\,,\qquad
\ea
where we have ignored operators that are either topological or removable by field redefinitions (and hence do not contribute independently to the amplitude) as well as dim-8 or higher operators that are irrelevant to this discussion. In practise the value of  $\b3 $ is dominated by the effects of the electron loops \cite{Drummond:1979pp},
\ba
\label{eq:b3ele}
\b3 =-\frac{\alpha}{360\pi m_{{\rm e}}^2}\(1+\mathcal{O}\(\frac{m_{\rm e}^2}{m_W^2}, \frac{m_{\rm e}^2}{m_{\rm meson}^2}\)\)\,,
\ea
where $m_e$ is the electron mass, $\alpha = q_e^2/(4\pi)$ is the fine-structure constant and $q_e$ is the electric charge.
Following the conventions laid out in the supplementary material, the non-zero definite helicity  $Ah\to Ah$ amplitudes are given by
\be
\begin{aligned}
&\mpl^2\cA_{++\to++}=\mpl^2\cA_{--\to--}=-\frac{s^2}{t}+\mathcal O(s^3)\,,\\
&\mpl^2\cA_{++\to--}=\mpl^2\cA_{--\to++}=2 \b3  t^2 +\mathcal O(s^3)\,,\\
&\mpl^2\cA_{++\to-+}=\mpl^2\cA_{--\to+-}=2 \b3  su\,,
\end{aligned}
\ee
where in the amplitude $\mathcal A_{h_1,h_2\to h_3,h_4}$, $h_1,h_3$ are the photon polarisations and $h_2, h_4$ the graviton ones. These amplitudes are consistent with those derived in \cite{Bjerrum-Bohr:2014lea} when $\b3=0$.
The remaining amplitudes can be expressed in terms of the amplitudes given above by using the fact that the amplitudes are symmetric under parity and the $s,u$ crossing symmetry, i.e.
\be
\cA_{h_1,h_2\to h_3,h_4}(s,t,u)=\cA_{h_1,\bar h_4\to h_3,\bar h_2}(u,t,s)\,.
\ee
In particular this implies that $\cA_{+-\to+-}(s,t,u)=\cA_{++\to++}(u,t,s)$.
As one can see from the above results, there is no contribution to the positivity bounds on elastic definite-helicity amplitudes coming from the $\b3$ term.

We can further consider initial and final photon and graviton states with indefinite polarisations,
\ba
\left|{A}\right\rangle=a_+ \left|{+}\right\rangle+a_{-} \left|{-}\right\rangle\ \ {\rm and}\ \
\left|{h}\right\rangle=h_+\left|{+}\right\rangle+h_{-}\left|{-}\right\rangle\,,\ \nn
\ea
where $a_\pm$, $h_\pm$ are complex numbers normalised so that $|a_+|^2+|a_-|^2=1$, etc.
Subtracting the poles in all three channels, the result is sensitive to the indefinite state of the photon, (see supplementary material for details)
\be
\del_s^2 \hat \cA(0,0)=-8\b3 \text{Re}(a_+a_-^*)\,,
\ee
which is sign indefinite, regardless of the sign of $\b3$. In particular we may make the reasonable choice $a_+=1/\sqrt{2}$ and $a_- = {\rm Sign}(\b3)/\sqrt{2}$ for which
\be\label{finalamp}
\del_s^2 \hat \cA(0,0)=-4|\b3|\,.
\ee
This is where the application of the positivity bounds \eqref{bound1}  is particularly insightful as it leads to
\be
2 \beta_s  (\ln r_s)'+ 2 \beta_u  (\ln r_u)' > 4|\b3| +\beta_s (\ln \alpha'_s)'+\beta_u (\ln \alpha'_u)' \, .\nn
\ee
The gravitational positivity bounds are thus violated unless either $(\ln r)'$ or the Regge slope $(\ln \alpha')'$ are bounded by the ratio of the mass of the electron to its charge.
In this sense, the positivity bound can only be viewed as an IR constraint on UV physics.\\

\textit{\textbf{Reverse Bootstrapping.---}}
The  main observation is that the expected positivity bounds informed by a UV Regge behaviour \eqref{bound1} appear to be violated by the amount \eqref{finalamp} which is sensitive to the mass to charge ratio of the lightest charged particle in the SM, namely the electron. We will go through a list of potential implications and emphasize that irrespectively to how nature resolves this tension, the SM does provide a remarkable constraint on UV physics:
\begin{enumerate}
\item {\bf Regge residue:} The first possible resolution
    is that the residue of the Regge pole associated with the scattering of gravitons and photons varies at a scale related to the electron mass to charge ratio, $(\ln r)'\ge (m_e/q_e)^{-2}\sim \(10^{-3}{\rm GeV}\)^{-2}$ \footnote{One could naively have expected the scale associated with the residue derivative to be of order of the string scale, so a bound of $(\ln r)'\ge (m_e/q_e)^{-2}$ would correspond to a enhancement of those subleading effects by close to 40 orders of magnitude.}.
      This is a remarkable outcome as the Regge behaviour as indicated in \eqref{eq:Regge} is typically only related to the behaviour of UV physics and one would not expect it to be set by the electron mass scale.
\item {\bf Regge slope:}  Another way out could be  to set the scale of the Regge slope to be of order of the electron mass to charge ratio, $|\ln \alpha'|' \sim |\b3|^{-1}\sim m_e^2/q_e^2$.  This would then imply the presence  of a higher spin Regge pole already at the scale $\sqrt{m_e M_s/q_e}$ where $M_s=1/\sqrt{\alpha'}$, leading to an infinite tower of  higher spins starting at or below about $10^4$TeV,  \footnote{If the Regge slope satisfies $|\ln \alpha'|' \sim |\b3|^{-1}\sim m_e^2/q_e^2$, this would imply $\alpha(t) = 2+ \alpha' t + c \frac{q_e^2}{m_e^2}\alpha' t^2+\dots$ and  we would then have a spin-3 pole $\alpha(t)=3$ at a scale $\sqrt{t} \sim \sqrt{m_e M_s/q_e}\lesssim 10^{4}$TeV and a similarly or even more closely spaced infinite tower of higher spins thereafter.}.
\item {\bf Causality/Locality:} In order to derive the positivity bounds, a certain level of causality/locality has been postulated when assuming the Froissart-like bound \eqref{eq:Froissart}.   While this bound is preserved for amplitudes derived from perturbative string theory in $D\ge 5$, it is possible that it is not technically applicable in the context of gravitational EFTs. In $D=4$ the bound is known to be more subtle due to IR divergences, however by working with $\tilde \cA$ defined in \eqref{eq:Atilda} we have removed the dangerous graviton loops. If failure of this bound were the reason why the amplitude \eqref{finalamp} carries such a high level of negativity, the consequences for UV physics and string theory in particular would be significant.
\item {\bf Light loops \& gaplessness:}
    We have argued that to the order  we are interested in, the amplitude is insensitive to graviton and photon loops. Graviton loops are Planck scale suppressed and do not enter $\tilde \cA$ by construction. At low energies photon loops contribute at best as $s^2 t \log t$, so $\p_s^2 \hat \cA$ is finite at $s=t=0$ even if $\p_s^3 \hat \cA$ and higher derivatives of the amplitude are not.  More dangerous is the fact that photon loops may undermine the Froissart-like bound \eqref{eq:Froissart}.
    While technically possible as a resolution, it would be indicative of a nontrivial UV/IR mixing and fall under the previous category as a weakening of locality.

\end{enumerate}

The possibility of introducing other new physics is discussed in the supplementary material and we argue that other than the inclusion of an   infinite tower of higher spin at the TeV scale or lower, there is no new beyond standard model (BSM) physics nor non-minimal couplings that could ameliorate the situation.

For pragmatic reasons, we have focused our discussion on graviton--photon exchange as it shows a clear level of negativity within known SM physics. All the arguments presented here are however generic and apply to any $U(1)$. In particular for any other dark sector $U(1)$ or BSM, the pole-subtracted graviton--gauge field indefinite scattering amplitude will always acquire a negative contribution that scales as the mass of the lightest particle charged under this $U(1)$. For instance imagining a dark photon and  charged dark matter particles under this dark $U(1)$ as in \cite{Pospelov:2008zw}, one would expect the residue of the Regge behaviour to carry a scale as small as the lightest charged dark matter particle, a scale which could in principle be extremely low. In some of these models, the dark photon could also be massive hence avoiding any IR divergences issues. Whether we are dealing with the actual photon or with another gauge field, there are no other operators one could include into the EFT that would change our results and the positivity bounds cannot be read as a constraint on the cutoff of the EFT. Rather the constraint has to be imposed directly at the level of either the Regge behaviour, the Froissart bound, the mixing with IR loops or the presence of an infinite tower of higher spin states at the scale $\sqrt{m M_s/q}$, where $m$ and $q$ are the mass and charge of the lightest charged particle. Interestingly, the scale associated with this behaviour is closely related to that entering the Weak Gravity Conjecture \cite{Cheung:2014ega,Andriolo:2018lvp,Hamada:2018dde,Aalsma:2020duv,Alberte:2020bdz}.\\
Irrespectively on which avenue is the most likely explanation, our findings show how SM physics has to be woven into UV physics.\\


\begin{acknowledgments}
The work of AJT and CdR is supported by STFC grants ST/P000762/1 and ST/T000791/1. CdR thanks the Royal Society for support at ICL through a Wolfson Research Merit Award. LA and CdR are supported by the European Union Horizon 2020 Research Council grant 724659 MassiveCosmo ERC2016COG. CdR is also supported by a Simons Foundation award ID 555326 under the Simons Foundation Origins of the Universe initiative, Cosmology Beyond Einstein's Theory and by a Simons Investigator award 690508. SJ is supported by an STFC studentship. AJT thanks the Royal Society for support at ICL through a Wolfson Research Merit Award.
\end{acknowledgments}

\appendix
\section{Supplementary Material}
\textit{\textbf{Conventions.---}}
Throughout the manuscript, we use mostly minus signature $(+,-,-,-)$, define the Planck mass as $\mpl^2\equiv(8\pi G)^{-1}$ and use conventions where  $
R^{\rho}\,_{\mu\sigma\nu}=\del_{\sigma}\Gamma^{\rho}\,_{\mu\nu}+\dots$.

Consider a scattering process $A(h_1,k_1)\,h(h_2,k_2)\to A(h_3,k_3)\,h(h_4,k_4)$ where $h_1, h_3$ and  $h_2,h_4$ are the helicities of ingoing and outgoing photons and gravitons respectively and $k_i^\mu$ are their four-momenta. We parameterise the latter as
\be
k^\mu_i=k(1,\sin\theta_i,0,\cos\theta_i)\,
\ee
with $\theta_1=0\,,\theta_2=\pi\,,\theta_3=\theta\,,\theta_4=\pi+\theta$. We write the photon polarisation vectors in helicity basis as
\be
\varepsilon^\mu (h_i,\theta_i)=\frac{1}{\sqrt{2}}(0,\cos\theta_i,ih_i,-\sin\theta_i)
\ee
with $h_i=\{h_1,h_3\}$ taking the values $h_i=\pm 1$ corresponding to helicity $\pm 1$ photon polarisations. Similarly the graviton polarisations are written as
\be
\varepsilon^{\mu\nu}(h_i,\theta_i)=\varepsilon^\mu (h_i,\theta_i)\varepsilon^\nu(h_i,\theta_i)\,
\ee
with $h_i=\{h_2,h_4\}$ taking the values $h_i=\pm 1$ corresponding to helicity $\pm 2$ graviton polarisations.\\

\textit{\textbf{Indefinite hAhA amplitude.---}}
Consider initial and final photon and graviton states with indefinite polarisations, we can take
\ba
\left|{A}\right\rangle=a_+ \left|{+}\right\rangle+a_{-} \left|{-}\right\rangle\ \ {\rm and}\ \
\left|{h}\right\rangle=h_+\left|{+}\right\rangle+h_{-}\left|{-}\right\rangle\,,\ \nn
\ea
where $a_\pm$, $h_\pm$ are complex numbers normalised so that $|a_+|^2+|a_-|^2=1$, and $|h_+|^2+|h_-|^2=1$.
 This gives the initial state,
\be
\ket{i}=\ket{A}\otimes\ket{h}=\alpha_1 \ket{++}+\alpha_2\ket{+-}+\alpha_3\ket{-+}+\alpha_4\ket{--}\,,\nn
\ee
with $\alpha_1=a_{+}h_{+},\,\alpha_2=a_{+}h_{-},\,\alpha_3=a_{-}h_{+},\,\alpha_4=a_{-}h_{-}$.
Positivity bounds can only be applied to elastic scattering amplitudes with $\left\langle{f}\right|=(\left|{i}\right\rangle)^{\dagger}$\,. Given the above expression for the initial state it is clear that the indefinite elastic amplitude is a linear combination of elastic and inelastic definite-polarisation amplitudes and so may lead to some non-zero $\b3$-dependence. Indeed, the elastic amplitude is,
\be
\begin{aligned}
\bra{i}\hat T\ket{i}&=(|\alpha_1|^2+|\alpha_4|^2)\bra{++}\hat T\ket{++}\\
&+(\alpha_1\alpha_4^*+\alpha_4\alpha_1^*)\bra{--}\hat T\ket{++}\\
&+(|\alpha_2|^2+|\alpha_3|^2)\bra{+-}\hat T\ket{+-}\\
&+(\alpha_1\alpha_3^*+\alpha_4\alpha_2^*)\bra{-+}\hat T\ket{++}\\
&+(\alpha_2\alpha_4^*+\alpha_3\alpha_1^*)\bra{--}\hat T\ket{+-}\,,\\
\end{aligned}
\ee
where we have used the parity symmetry and some of the results for the definite-helicity amplitudes given above. Subtracting the poles in all three channels, the result ends up only depending on the  indefinite state of the photon and not on that of the graviton,
\be
\del_s^2 \hat \cA(0,0)=-8\b3 \text{Re}(a_+a_-^*)\,.
\ee

\textit{\textbf{New Physics?---}} At the level of \eqref{EH},  there are no other {\it local} and {\it covariant} operators that would affect our final result \eqref{finalamp}. The only way to change our amplitude \eqref{finalamp} would be to advocate the presence of an additional bare coupling between the graviton and the photon already active within the SM \eqref{SM}, $\bt3 {\rm Riem}F^2$ tuned so as to precisely cancel the effect of SM loops, namely $\bt3\equiv -\b3+\mathcal{O}(\Lambda^{-2})$, with $\Lambda\gtrsim$ TeV a new cutoff scale, so that what enters in \eqref{finalamp} at low-energy is $\b3^{\rm total}=\b3+\bt3\sim \mathcal{O}(\Lambda^{-2})$.

From a theoretical point of view, such a tuning could in principle be advocated and could for instance follow from integrating out charged BSM particles. However in practice, knowing that charged BSM particles can at best arise at TeV scale, it would require at a minimum about $10^{12}$ BSM particles with masses, charges and spins precisely tuned so that their loops leads to an effective  $\bt3\equiv -\b3+\mathcal{O}({\rm TeV}^{-2})$. Even so, the inclusion of such a non-minimal coupling at a scale $\bt3\sim m_e^{-2}$ would not help either. One could then apply improved bounds at energy scales above $m_e$ but below $\Lambda$, where known SM loops are subtracted out following the same procedure as in \cite{Alberte:2020bdz}. The resulting improved positivity bounds would then involve solely $\bt3$ and not $\b3+\bt3$. Following the same argument as previously, there will always be an indefinite scattering for which $\del_s^2 \hat \cA^{({\rm improved})}(0,0)\sim-4|\bt3|$. Aside from a low-scale  infinite tower of higher spin, there is therefore no new BSM physics nor non-minimal couplings that could ameliorate the situation.\\

\onecolumngrid

\twocolumngrid

\bibliography{references}

\end{document}